# Non Standard Interactions in Hyperons Decays with di-Neutrinos in the final state

Abrar Ahmed*;Shakeel Mahmood†; Farida Tahir‡; Maimoona Razzaq§

Physics Department, Air University, PAF Complex, Service Road E-9, Islamabad, Pakistan *

Physics Department, COMSATS Institute of Information Technology, Islamabad, Pakistan†

**Abstract**

Rare decays of hyperons involving di-neutrinos in the final state $\Lambda \longrightarrow n\overline{\nu}\nu,$, $\Sigma^{+} \longrightarrow p\overline{\nu}\nu$, $\Xi^{0} \longrightarrow \Lambda\overline{\nu}\nu$, $\Xi^{0} \longrightarrow \Sigma^{0}\overline{\nu}\nu$, $\Xi^{-} \longrightarrow \Sigma^{-}\overline{\nu}\nu$, $\Omega^{-} \longrightarrow \Xi^{-}\overline{\nu}\nu$ in the standard model and beyond are studied. It is claimed that the branching ratios (Brs) calculated in this work are 2 times the values of the 331 model and exceptionally larger than the previously calculated values. The Br of these decays are not in the access of any existing experiment but new physics (NP) can enhance Br considerably. For the search of NP the nonstandard neutrino interactions (NSI) are explored and constrains on NSI free parameters $\epsilon^{uL}_{\alpha\beta}$ are found. Stringent bounds on $\epsilon^{uL}_{\tau\beta}$ $(\beta = e, \mu, \tau)$ of O $(10^{-2})$ are obtained. It is shown that Brs could be in the range of BES III, if the constraints are O(0.3).

Keywords: NSIS; New Physics; Rare Decays; FCNC

PACS numbers: 12.60.-i, 13.15.+g, 13.20.-v

## 1 Introduction

Standard Model (SM) predictions have been verified experimentally to the highest level of precision [1]. However, along with theoretical inconsistency, it lacks any explanation for a possible pattern for particle masses. The experiments on B mesons [2] are also giving some cracks in SM. We need New Physics (NP) to explain dark matter and matter anti-matter asymmetry. Gravity is not included in the SM. Theoretically SM has limitations and there can be some new particles as well as new interactions. But, so far, the only concrete evidence against SM has been provided by the neutrino oscillations [3, 4]. It has been believed that SM is a low energy approximation of more general theory. So, many theoretical extensions of SM have been presented. To explore NP the study of mesonic rare decays involving neutrinos in final state two processes $K^+ \to \pi^+\overline{\nu}\nu$ and $K_L^0 \to \pi^0\overline{\nu}\nu$ has been used due to their theoretical cleanness. These decays proceed through flavor changing neutral currents (FCNC) which are highly suppressed [5] in SM. Suppression at tree level due to GIM mechanism [6] and occurrence at the loop level is setting their Br to a very small value [7]. The discrepancies between experiments and theory SM for such reactions provide us an excellent window towards NP. New particles can be added in the loops to improve the theory. So, FCNC reactions involving neutrinos in the final state can be interesting. $K^+ \to \pi^+\overline{\nu}\nu$ has been used for the search of NP but there is a very small difference between theory and experiment for this reaction. Due to the involvement of neutrinos in the final state of this reaction NSI can be invoked easily [8]. The effect of NSI in rare decays of mesons has been studied in [8, 9, 10, 11, 12, 13, 14, 15]. Similar to this mesonic rare decay we have hyperons rare decays ($\Lambda \longrightarrow n\overline{\nu}\nu$, $\Sigma^+ \longrightarrow p\overline{\nu}\nu$, $\Xi^0 \longrightarrow \Lambda\overline{\nu}\nu$, $\Xi^0 \longrightarrow \Sigma^0\overline{\nu}\nu, \Xi^- \longrightarrow \Sigma^-\overline{\nu}\nu$ and $\Omega^- \longrightarrow \Xi^-\overline{\nu}\nu$), which can be observed in BES-III. These decays can be written as

$$B_{initial} \longrightarrow B_{final}\overline{\nu}_\alpha\nu_\beta$$

where mass of $B_{initial}$ > mass of $B_{final}$, and $B_{initial}$, $B_{final}$ represent lightest hyperons and baryons respectively.

*smart5733@gmail.com

†Corresponding author:*shakeel_ mahmood@hotmail.com;shakeel.mahmood@mail.au.edu.pk*

‡farida_tahir@comsats.edu.pk

§maimoonarazzaq10@gmail.com

At the quark level all these reactions are represented by the equation

$$s \to d\nu_\alpha\overline{\nu}_\beta$$

where $s$ and $d$ are representing strange and down quarks having $\alpha = \beta$, allowed in SM. The effective Lagrangian can be written as:

$$L_{eff}^{SM} = -2\sqrt{2}G_F(\overline{\nu}_\alpha\gamma_\mu L\nu_\alpha)(\overline{f}\gamma^\mu Pf)$$

where $\alpha$ corresponds to the light neutrino flavor, $f$ denotes a charged lepton or quark, where we are only dealing with quarks and $P = R$ or $L$ with $R(L) = (\frac{1\pm\gamma_5}{2})$.

The case $\alpha \neq \beta$ is strictly forbidden in SM, only possible in new physics scenario for which we are using non standard interactions (NSI).

NP can be searched in two ways: *a*) Model independent way and *b*) Model Dependent way, e.g., left right symmetric model, 331 model, SUSY etc. But, so far, all the searches for model dependent particles or interaction are providing negative results, leaving no choice but to adopt model independent approach such as NSI. We study the decays $s \longrightarrow d\ \nu\overline{\nu}$ to find out the contribution from the NP. We concentrate only on $\Lambda \longrightarrow n\overline{\nu}\nu$, $\Sigma^+ \longrightarrow p\overline{\nu}\nu$, $\Xi^0 \longrightarrow \Lambda\overline{\nu}\nu$, $\Xi^0 \longrightarrow \Sigma^0\overline{\nu}\nu, \Xi^- \longrightarrow \Sigma^-\overline{\nu}\nu, \Omega^- \longrightarrow \Xi^-\overline{\nu}\nu$.

This article is arranged as follows. After introduction In Section II we give experimental status of selected rare decays and then theory is discussed. Sections III and IV are devoted for the expressions from which branching ratios in both the SM and NSI can be calculated. The comparison is made from plots. Finally results and conclusions are provided in Section V.

## 2 Experimental Prospects and Effective Hamiltonian

These processes are yet to detect in the experiments but Beijing Electron Spectrometer (BES III) is proposed to observe $J/\Psi$ decay into hyperon pairs. It is estimated that about $10^6$ to$10^{10}$ hyperons will be produced. This would be an excellent opportunity to observe $\Lambda$, $\Sigma$, $\Xi$ and $\Omega$. This experiment is capable of observing Br of $10^{-5}$ to $10^{-8}$of these decays. The expected results are published in [33] and listed in Table I.

In the study of rare decays the Effective Hamiltonian (**EH**) [16, 17] is obtained by the use of operator product expansion and renormalization group. This approach can easily separate short-distance contributions from long distance contributions. Short distance contribution can be studied by using perturbation theory. The long distance contributions are carried by matrix elements of the operators. These matrix elements require non-perturbative methods for their calculation and so they are model dependent. $s \longrightarrow d\ \nu\overline{\nu}$ is a short distance dominated process and its hadronic part can be extracted by using a tree level process, making it a clean process. The effective Lagrangian for such interactions in model independent way is given in [18] and can be written as

$$L_{eff}^{NSI} = -2\sqrt{2}G_F\left[\sum_{\alpha=\beta}\epsilon_{\alpha\beta}^{fP}(\overline{\nu}_\alpha\gamma_\mu L\nu_\beta)(\overline{f}\gamma^\mu Pf)+\sum_{\alpha\neq\beta}\epsilon_{\alpha\beta}^{fP}(\overline{\nu}_\alpha\gamma_\mu L\nu_\beta)(\overline{f}\gamma^\mu Pf)\right], \tag{1}$$

where $\epsilon_{\alpha\beta}^{fP}$ is the parameter for NSI, which carries information about dynamics. NSI are thought to be well-matched with the oscillation effects along with new features in neutrino searches [19, 20, 21, 22, 23]. Expected constraints for first two generations of leptons $\epsilon_{\tau l}^{fP}$ ($l = e,\mu$ ) by tree level processes are at $O(10^{-3})$ by future $\sin^2(\theta)$ experiments [24]. For third generation ($\tau$) we need decays which occur at loop level whose current limit is $O(1)$. The limit of $O(0.3)$ is expected for third generation ($\tau$) from KamLAND data [25] and solar neutrino data [26, 27] .

The Standard Model interactions and nonstandard interactions are shown by diagrams as given in Figure1 and 2 respectively, just like Figure 1(b) we have a third diagram where the role of $W$ and up type-quarks in the loop is interchanged. The effective Hamiltonian is shown of the process $s \longrightarrow d\ \overline{v}v$ in SM and NSI is given by

$$H_{eff}^{SM} = \frac{G_F}{\sqrt{2}}\frac{\alpha_{em}}{2\pi\sin^2\theta_W}\sum_{\alpha,\beta=e,\mu,\tau}(V_{cd}^*V_{cs}X_{NL}^l + V_{td}^*V_{ts}X(x_t))(\overline{s}d)_{V-A}(\nu_\alpha\overline{\nu}_\alpha)_{V-A} \tag{2}$$

where $X_{NL}^l$ is the next to leading order contribution of the charm quark and similarly $X(x_t)$ is the top quark contributions [17]. In this case NSI becomes

$$H_{eff}^{NSI} = \frac{G_F}{\sqrt{2}}(V_{us}^*V_{ud}\frac{\alpha_{em}}{4\pi\sin^2(\theta_W)}\epsilon_{\alpha\beta}^{uL}\ln\frac{\Lambda}{m_W})(\nu_\alpha\overline{\nu}_\beta)_{V-A}(\overline{s}d)_{V-A}, \tag{3}$$

where $\Lambda$ represents energy scale for new physics and it is above the electroweak scale. As new physics can differentiate between different flavors of neutrino, so we have no summation, contrary to SM.

## 3 Branching Ratios of Decays in SM, NSI and 331 Model

We get the branching ratios for such reactions by normalizing with a tree level process which are by isospin symmetry which will reduce hadronic uncertainties and experimental Br of these tree level processes are given in table 1. Factor of three is due to summation over three flavours of neutrinos.

$$\begin{aligned}\frac{Br(\Lambda\to n\nu\overline{\nu})_{SM}}{Br(\Lambda\to pe^{-}\overline{\nu}_e)} = \frac{Br(\Sigma^{+}\to p\nu\overline{\nu})_{SM}}{Br(\Sigma^{-}\to ne^{-}\overline{\nu})} &= \frac{Br(\Xi\to\Lambda\nu\overline{\nu})_{SM}}{Br(\Xi^{-}\to\Lambda e^{-}\overline{\nu}_e)} \\ &= \frac{Br(\Xi^{0}\to\Sigma^{0}\nu\overline{\nu})_{SM}}{Br(\Xi^{0}\to\Sigma^{+}e^{-}\overline{\nu}_e)} = \frac{Br(\Xi^{-}\to\Sigma^{-}\nu\overline{\nu})_{SM}}{Br(\Xi^{-}\to\Sigma^{0}e^{-}\overline{\nu}_e)} \\ &= \frac{Br(\Omega^{-}\to\Xi^{-}\nu\overline{\nu})_{SM}}{Br(\Omega^{-}\to\Xi^{0}e^{-}\overline{\nu}_e)} = A \end{aligned} \quad (4)$$

Provided that $r_{iso}$ denote the isospin breaking effect similar to $r_{K^{+}}=0.901$, for which

$$\langle\pi^{+}\left|(\overline{s}d)_{V-A}\right|K^{+}\rangle=\sqrt{2}\langle\pi^{0}\left|(\overline{s}u)_{V-A}\right|K^{+}\rangle$$

and $A=\frac{3\alpha_{em}^{2}r_{iso}}{|V_{us}|^{2}2\pi^{2}\sin^{4}\theta_{W}}[(\frac{\mathrm{Im}\,\lambda_t}{\lambda^5}X(x_t))^2+(\frac{\mathrm{Re}\,\lambda_t}{\lambda^5}+\frac{\mathrm{Re}\,\lambda_c}{\lambda}(P_c(X)+\delta P_{c,u}))^2]$ is given and calculated in [28] for $k^{+}\to\pi^{+}\nu\overline{\nu}$ and $\lambda_i=V_{is}V_{id}$ $\lambda_5=V_{us}=0.2252$; $V_{ud}=0.97425$; $\theta_w=28.7^{\circ}$, $P_c(X)=0.369$, $\delta P_{c,u}=0.04$ [29]. Factor $A$ remains the same for our reactions and these are linked with same isospin symmetry, so we are using the same values.

Before proceeding forward, let's look at $k^{+}\to\pi^{+}\nu\overline{\nu}$ whose experimental value is available and Br ratio corresponding to nonstandard interactions is calculated in [30] as,

$$Br(k^{+}\to\pi^{+}\nu\overline{\nu})_{NSI}=Br(k^{+}\to\pi^{+}\nu\overline{\nu})_{\exp rimental}-Br(k^{+}\to\pi^{+}\nu\overline{\nu})_{SM}=O(10^{-11})$$

$$Br(k^{+}\to\pi^{+}\nu\overline{\nu})_{NSI}=\frac{\alpha_{em}^{2}r_{iso}}{|V_{us}|^{2}2\pi^{2}\sin^{4}\theta_{w}}|V_{us}^{*}V_{ud}\frac{1}{2}\epsilon_{\alpha\beta}^{uL}\ln\frac{\Lambda}{m_W}|^{2}Br(k^{+}\to\pi^{0}e^{+}\nu_e)$$

with $\epsilon_{\alpha\beta}^{uL}\leq\frac{8.8\times10^{-3}}{\ln\frac{\Lambda}{m_W}}$. As $\alpha$ and $\beta$ can be from any leptons family, we take tau neutrions, $\epsilon_{\tau\tau}^{uL}$ is $O(10^{-2})$

NSI can differentiate between neutrino flavours, so no summation. NSI Brs are calculated by using the Hamiltonian of Eq. (3) and given by the following expressions

$$Br(\Lambda\to n\nu\overline{\nu})_{NSI}=\frac{\alpha_{em}^{2}r_{iso}}{|V_{us}|^{2}2\pi^{2}\sin^{4}\theta_{W}}|V_{us}^{*}V_{ud}\frac{1}{2}\epsilon_{\alpha\beta}^{uL}\ln\frac{\Lambda}{m_W}|^{2}Br(\Lambda\to pe^{-}\overline{\nu}_e)$$

The nonstandard branching ratio is equal to

$$Br(\Lambda\rightarrow n\nu\overline{\nu})_{NSI}=Br(\Lambda\rightarrow n\nu\overline{\nu})_{\exp rimental}-Br(\Lambda\rightarrow n\nu\overline{\nu})_{SM}$$

but due to unavailability of experimental value of, we use $\epsilon_{\alpha\beta}^{uL}\leq\frac{8.8\times10^{-3}}{\ln\frac{\Lambda}{m_{W}}}$ and it becomes 0.01 at $\Lambda=4m_{W}$

Similarly BSIs branching ratios can be found for other decays

$$\begin{array}{l}
Br(\Sigma^{+}\rightarrow p\nu\overline{\nu})_{NSI}=\frac{\alpha_{em}^{2}r_{iso}}{|V_{us}|^{2}2\pi\sin n^{4}\theta_{W}}|V_{us}^{*}V_{ud}\frac{1}{2}\epsilon_{\alpha\beta}^{uL}\ln\frac{\Lambda}{m_{W}}|^{2}Br(\Sigma^{-}\rightarrow ne^{-}\overline{\nu}),\\
Br(\Xi\rightarrow\Lambda\nu\overline{\nu})_{NSI}=\frac{\alpha_{em}^{2}r_{iso}}{|V_{us}|^{2}2\pi\sin n^{4}\theta_{W}}|V_{us}^{*}V_{ud}\frac{1}{2}\epsilon_{\alpha\beta}^{uL}\ln\frac{\Lambda}{m_{W}}|^{2}Br(\Xi^{-}\rightarrow\Lambda e^{-}\overline{\nu}),\\
Br(\Xi^{0}\rightarrow\Sigma^{0}\nu\overline{\nu})_{NSI}=\frac{\alpha_{em}^{2}r_{iso}}{|V_{us}|^{2}2\pi\sin n^{4}\theta_{W}}|V_{us}^{*}V_{ud}\frac{1}{2}\epsilon_{\alpha\beta}^{uL}\ln\frac{\Lambda}{m_{W}}|^{2}Br(\Xi^{0}\rightarrow\Sigma^{+}e^{-}\overline{\nu}_{e}),\\
Br(\Xi^{-}\rightarrow\Sigma^{-}\nu\overline{\nu})_{NSI}=\frac{\alpha_{em}^{2}r_{iso}}{|V_{us}|^{2}2\pi\sin n^{4}\theta_{W}}|V_{us}^{*}V_{ud}\frac{1}{2}\epsilon_{\alpha\beta}^{uL}\ln\frac{\Lambda}{m_{W}}|^{2}Br(\Xi^{-}\rightarrow\Sigma^{0}e^{-}\overline{\nu}_{e}),\\
Br(\Omega^{-}\rightarrow\Xi^{-}\nu\overline{\nu})_{NSI}=\frac{\alpha_{em}^{2}r_{iso}}{|V_{us}|^{2}2\pi\sin n^{4}\theta_{W}}|V_{us}^{*}V_{ud}\frac{1}{2}\epsilon_{\alpha\beta}^{uL}\ln\frac{\Lambda}{m_{W}}|^{2}Br(\Omega^{-}\rightarrow\Xi^{0}e^{-}\overline{\nu}_{e}).
\end{array} \tag{5}$$

As in $\epsilon_{\alpha\beta}^{UL}$, $\alpha$ and $\beta$ can take any leptons, we pick out $\tau$ (just like $k^{+}\rightarrow\pi^{+}\nu\overline{\nu}$) as one of the lepton whose chances are more in normal mass hierarchy of neutrinos due to decays of hyperons. The plots of Br and new physics parameter $\epsilon_{\alpha\beta}^{uL}$ are done at different values of new physics energy scale $\Lambda$. Figure 3 a is providing value of $\epsilon_{\tau\tau}$, $O(10^{-2})$ with which very stringent as compared with current experimental value (O(1). Similarly, Figure 4 is presenting a future expected value of $O(0.3)$. For other lepton flavors $\epsilon_{ll}<\epsilon_{\tau\tau}$, the values would be less than the values for $\tau$. The values of parameters as well as the nonstandard interactions branching ratios are given in Table 1.

Finally, let us examine the 331 model for the calculation of the branching ratios [32], which is an extension of the SM at TeV scale and introduces $Z^{/}$ boson with mass 1 TeV. The Hamiltonian for this model will be changed to

$$H_{eff}^{Z^{/}}=\underset{l=e,\mu,\tau}{\Sigma}\widetilde{V}_{32}^{*}V_{31}(\frac{M_{Z}}{M_{Z^{/}}})^{2}\cos^{2}(\theta_{W})(\overline{s}d)_{V-A}(\nu_{l}\overline{\nu}_{l})_{V-A}+h.c.$$

with 331 parameters $Re[(\widetilde{V}_{32}^{*}V_{31})]=9.2\times10^{-6}$ and $Im[(\widetilde{V}_{32}^{*}V_{31})]=4.8\times10^{-8}$. The effect of this will be only to change $X(x_{t})$ to

$$X(x_{t})=X^{SM}(x_{t})+\Delta X$$

where $\Delta X=\frac{\sin^{2}(\theta_{W})\cos^{2}(\theta_{W})}{\alpha}\frac{2\pi}{3}\frac{\widetilde{V}_{32}^{*}V_{31}}{V_{ts}^{*}V_{td}}(\frac{M_{Z}}{M_{Z^{/}}})^{2}$. So, this modification is not changing the Brs as desired and the order of magnitude remains the same.

# 4 Conclusions

Standard model branching ratios calculated and listed in Table I are two times the values of the 331- model in reference [32] and $10-10^{2}$ enhanced as compared to reference [31]. It is evident from the plots and Table 1 that NSI branching ratios are very close to 331 model values. The value of the parameter $\epsilon_{\tau\tau}^{uL}$ corresponding to these branching ratios is 0.01. For the future limit on the parameter $\epsilon_{\tau\tau}^{uL}$, $O(0.3)$ the nonstandard branching ratios could become vary high, $10^{-8}$ to $10^{-9}$, which are very closed to the sensitivity of BES-III. If these reactions would not be detected in BES III than more stringent bounds on new physics parameter would be imposed. If the situation remains exactly the same as that for $k^+ \to \pi^+ \nu\overline{\nu}$, then the bounds on NSI parameter are $O(10^{-2})$.

## References

[1] J. Erler and M. Schott, Progress in Particle and Nuclear Physics 106, 68–119 (2019).

[2] E. Gibney, nature 18307 (2015).

[3] A. B. McDonald, "Nobel Lecture: The Sudbury Neutrino Observatory: Observation of flavor change for solar neutrinos," Rev. Mod. Phys. 88 030502 (2016).

[4] A. Bellerive et al. [SNO Collaboration], "The Sudbury Neutrino Observatory," Nucl. Phys. B 908, 30 (2016).

[5] G. Buchalla, A. J. Buras,and M. E. Lautenbacher, Rev. Mod. Phys. 68, (1996).

[6] L. Maiani, arXiv:1303.6154v1 [hep-ph], (2013).

[7] M. K. Qaillard and B. W. Lee, PRD 10 (1974).

[8] C. H. Chen, C. Q. Geng and T. C. Yuan, PRD 75, 077301 (2007).

[9] S. Mahmood, F. Tahir and A. Mir, MPLA 30, 1550004 (2015).

[10] S. Mahmood, F. Tahir and A. Mir, IJMPA 30, 1550024 (2015).

[11] S. Mahmood, F. Tahir and A. Mir, IJMPA 30, 1550013 (2015).

[12] S. Mahmood, F. Tahir and A. Mir, IJMPA 30, 1550154 (2015).

[13] S. Mahmood, F. Tahir and A. Mir, IJMPE 24, 1571001(2015) .

[14] S. Mahmood, F. Tahir and A. Mir, MPLA 24,1850171 (2018).

[15] S. Mahmood, F. Tahir and A. Mir, PEPAN Lett. 15, 204-246 (2018).

[16] W. BUCHMLJLLER, D. WYLER Nucl. Phys. B 268, 621-653 (1986).

[17] A. J. Buras, arXiv:hep-ph/0101336v1 30 Jan 2001.

[18] C.Soumya and R. Mohanta, PRD 94, 053008 (2016).

[19] L. Wolfenstein, PRD 17 2369 (1978).

[20] S.P. Mikheev and A.Y. Smirnov, Sov. J. Nucl. Phys. 42, 913 (1985).

[21] M.C. Gonzalez-Garcia et al., PRL 82, 3202 (1999).

[22] M.M. Guzzo, H. Nunokawa, P.C. de Holanda and O.L.G. Peres, PRD 64, 097301 (2001).

[23] M.M. Guzzo, A. Masiero and S.T. Petcov, Phys. Lett. B 260, 154 (1991).

[24] S. Davidson et al., JHEP 03, 011 (2003).

[25] A.B. Balantekin et al., Phys. Lett. B 613, 61–66 (2005).

[26] Y. Farzan, M. Tortola, arXiv:1710.09360v2 [hep-ph] 28 Jan 2018.

[27] S. Verma and S. Bhardwaj, Adv. High Energy Phys. 2019, 8464535 (2019).

[28] G. Buchalla, A. J. Buras and M. E. Lautenbacher, Rev. Mod. Phys.68,1125-1144 (1996).

[29] M. Tanabashi et al. (PDG) PRD 98, 030001 (2018).

[30] Chuan-Hung Chen, Chao-Qiang Geng and Tzu-Chiang Yuan, PRD 75, 077301 (2007).

[31] T. Jasak , JHEP 04, 104 (2019).

[32] X. H. Hu and Z. X. Zhao, Chinese Physcs C 43, 093104 (2019).

[33] H. B. Li, front. Phys. 12 (5) 121301 (2017).

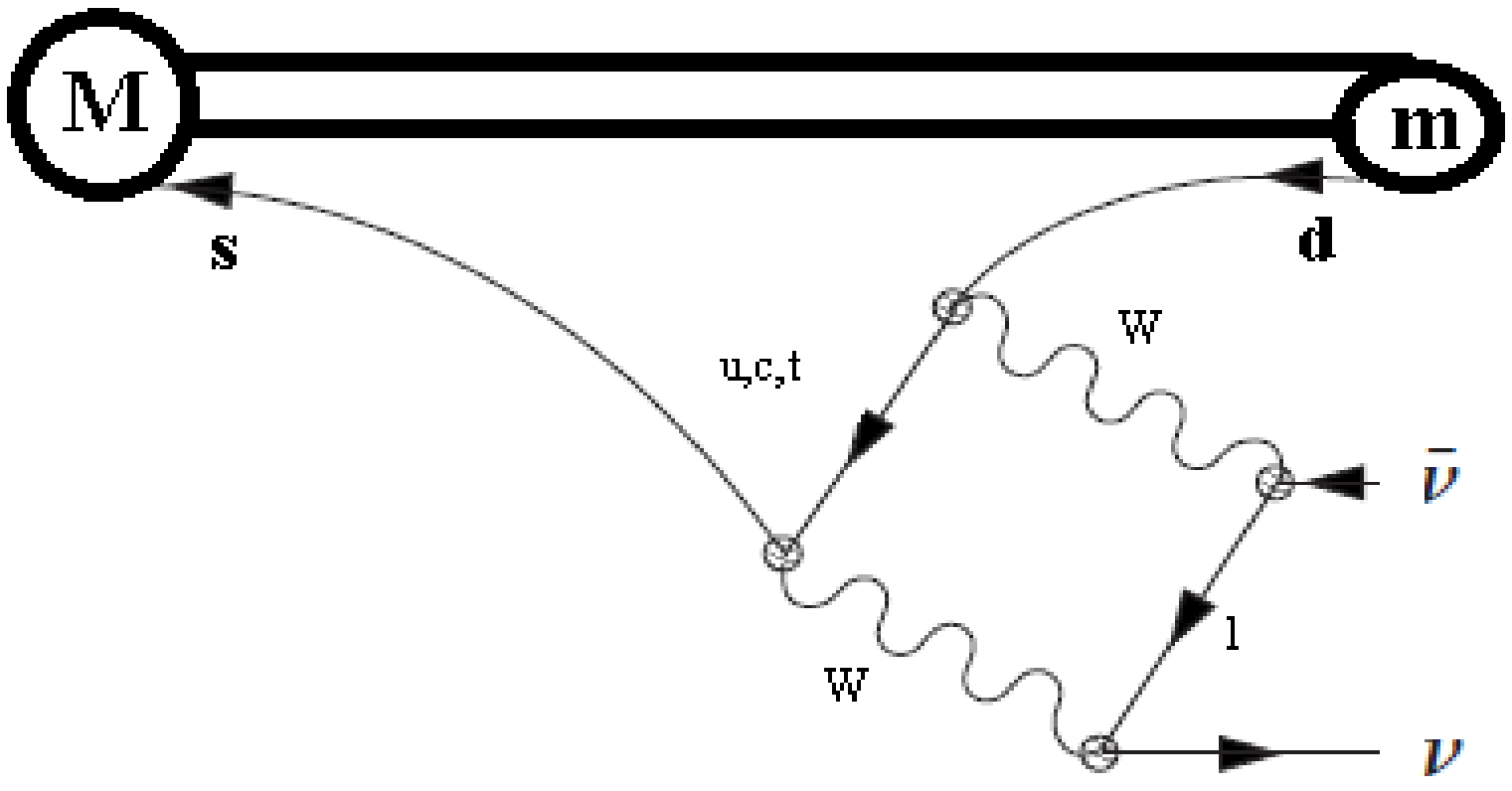

Figure 1 (a) BOX Diagram of Quark Level Process $s \to dv\overline{v}$

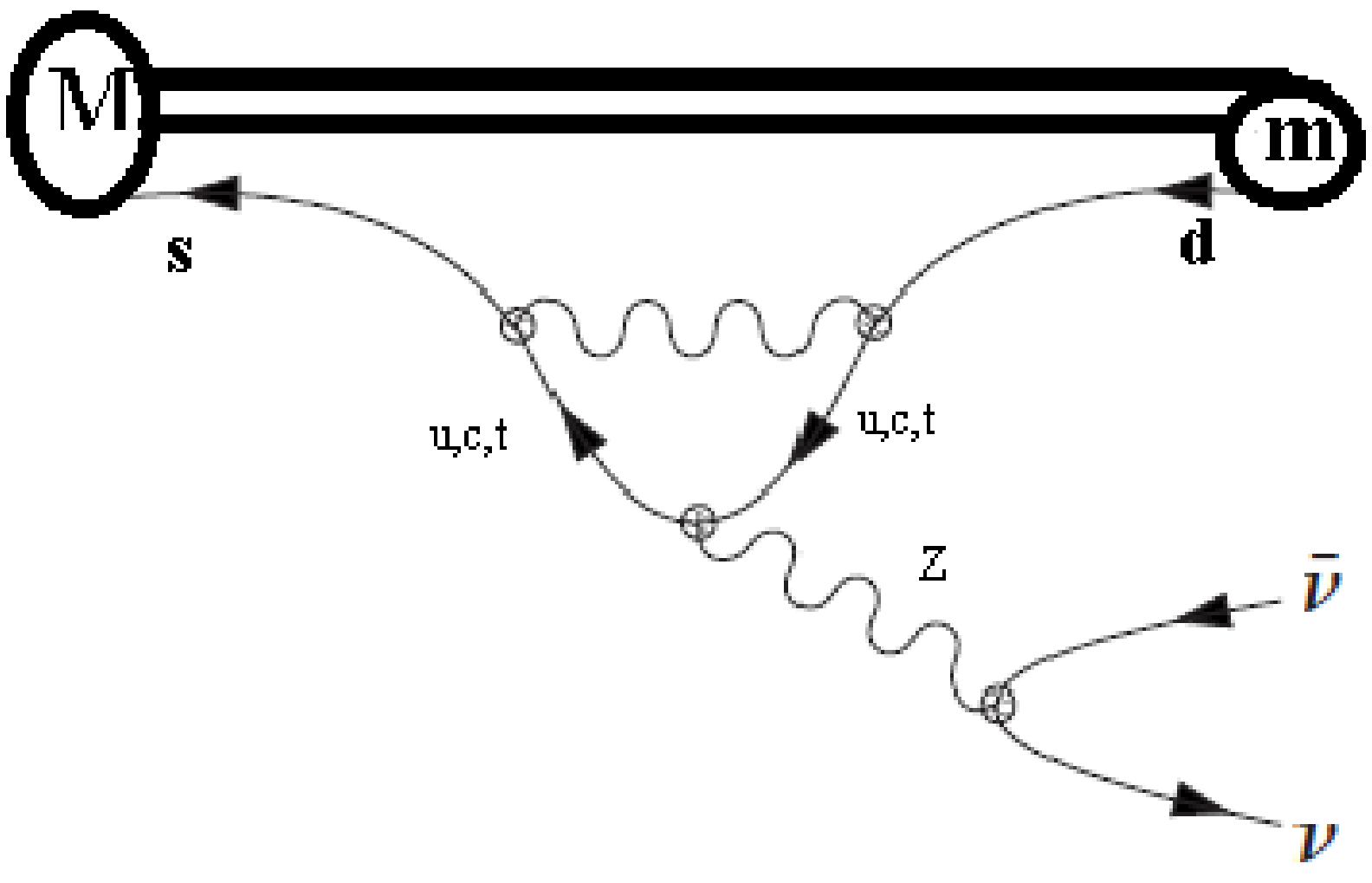

Figure 1(b) Penguin Diagram of Quark Level Process $s \to dv\overline{v}$

Figure 1: Figures 1 (a) and (b) are Standard Model Diagrames

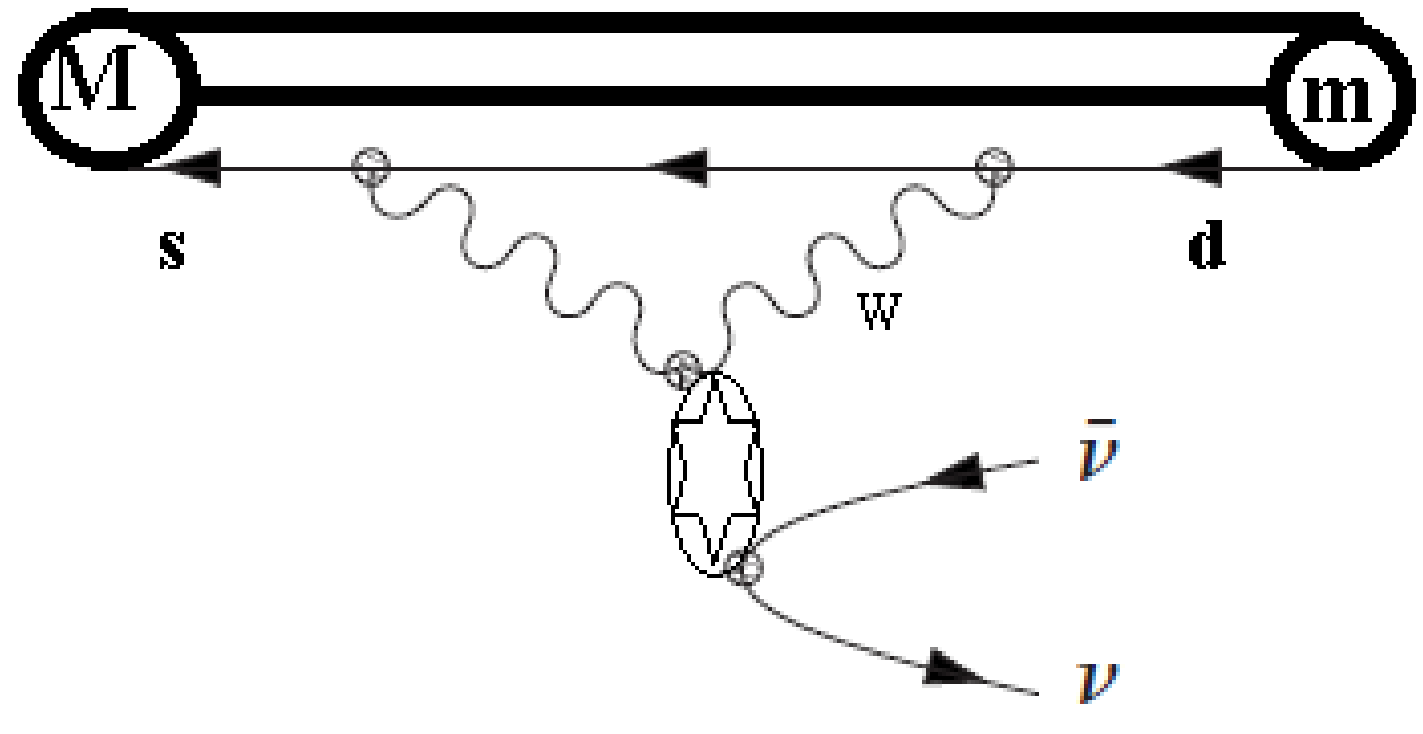

Figure 2 (a) Nonstand Penguin Diagram of $s \rightarrow dv\overline{v}$

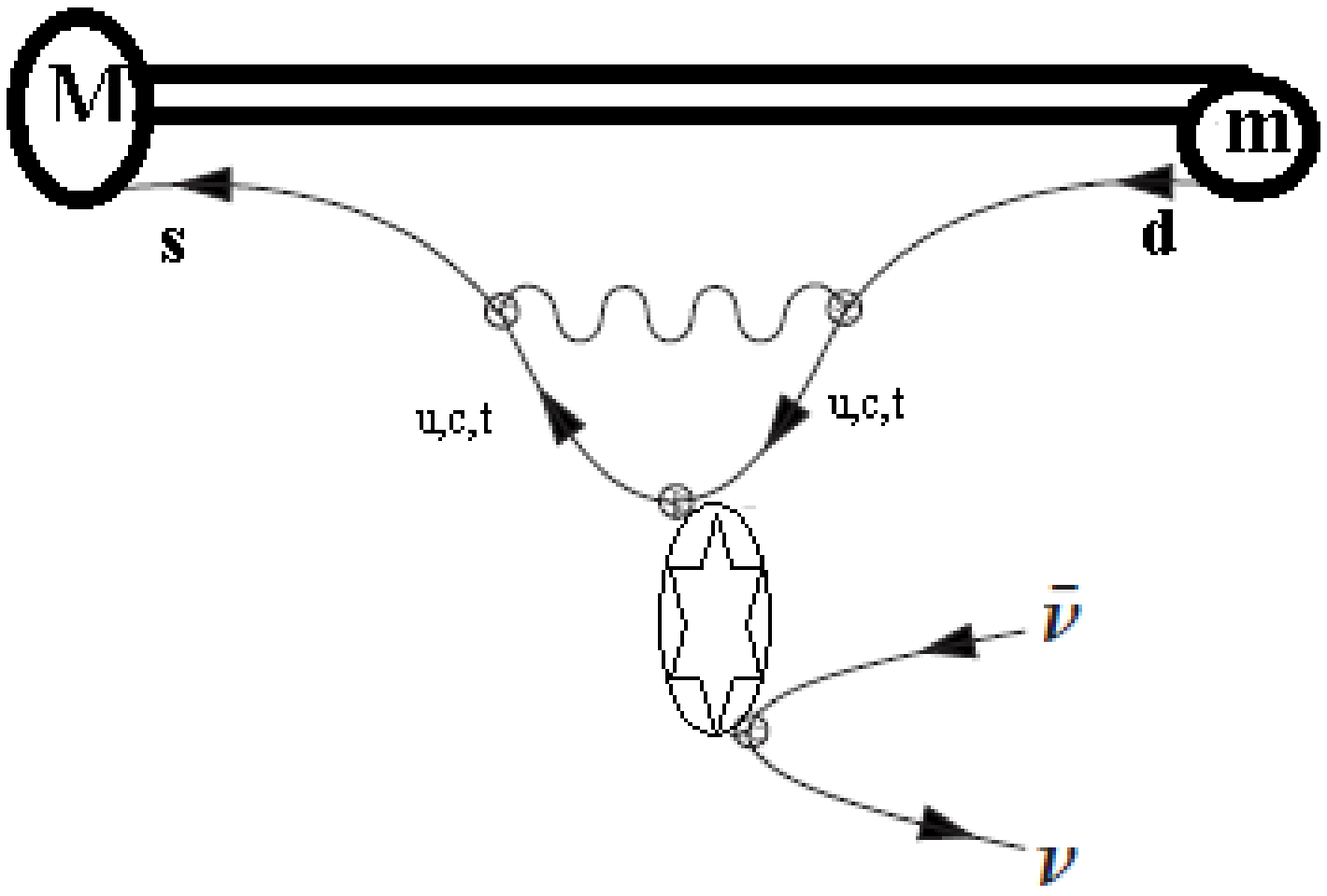

Figure 2 (b) Nonstand Penguin Diagram of $s \rightarrow dv\overline{v}$

Figure 2: Nonstandard Diagrams

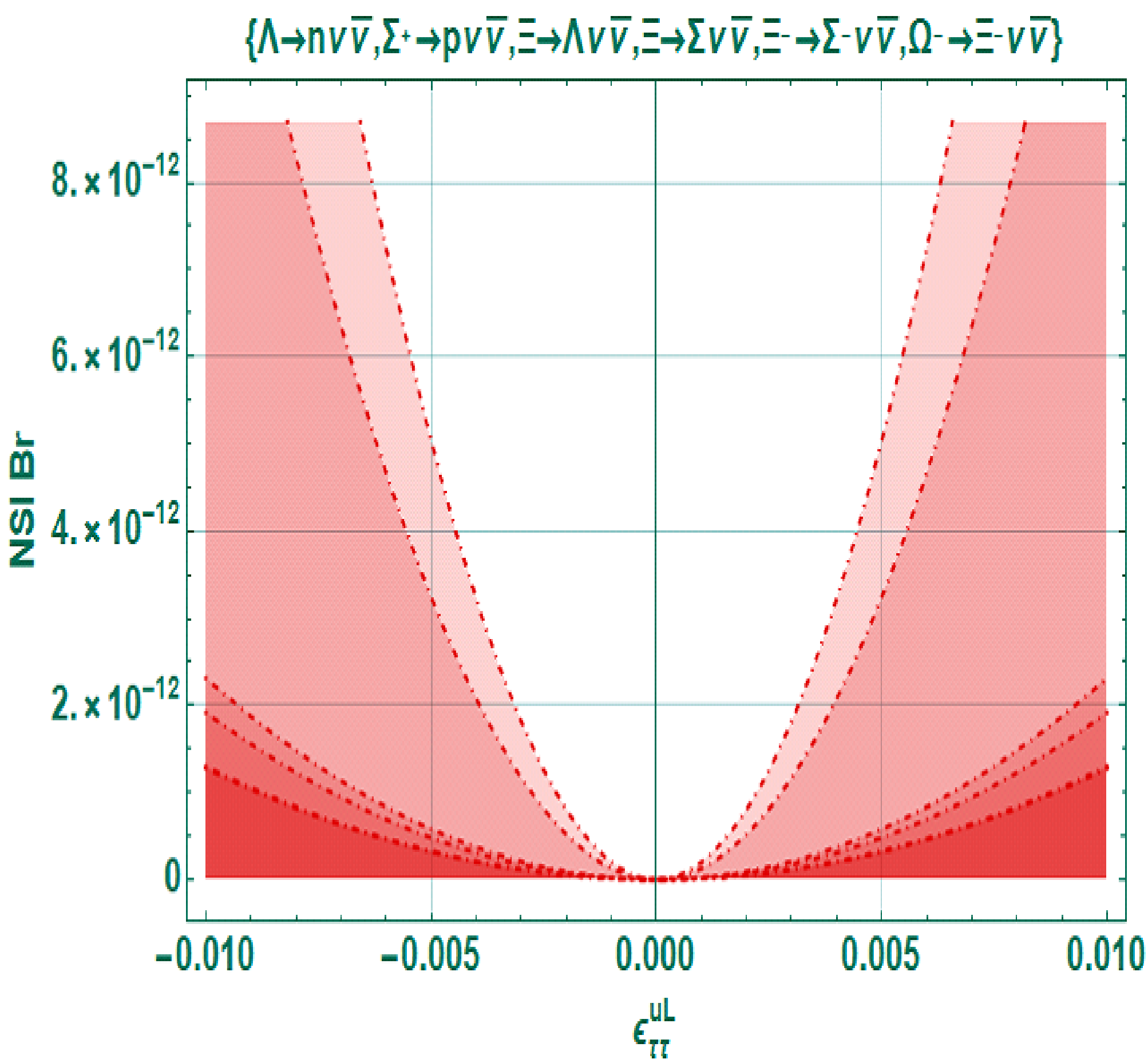

Combine graph of all process with New Physics Parameter $\epsilon_{\tau\tau}$, $O(10^{-2})$

Figure 3:

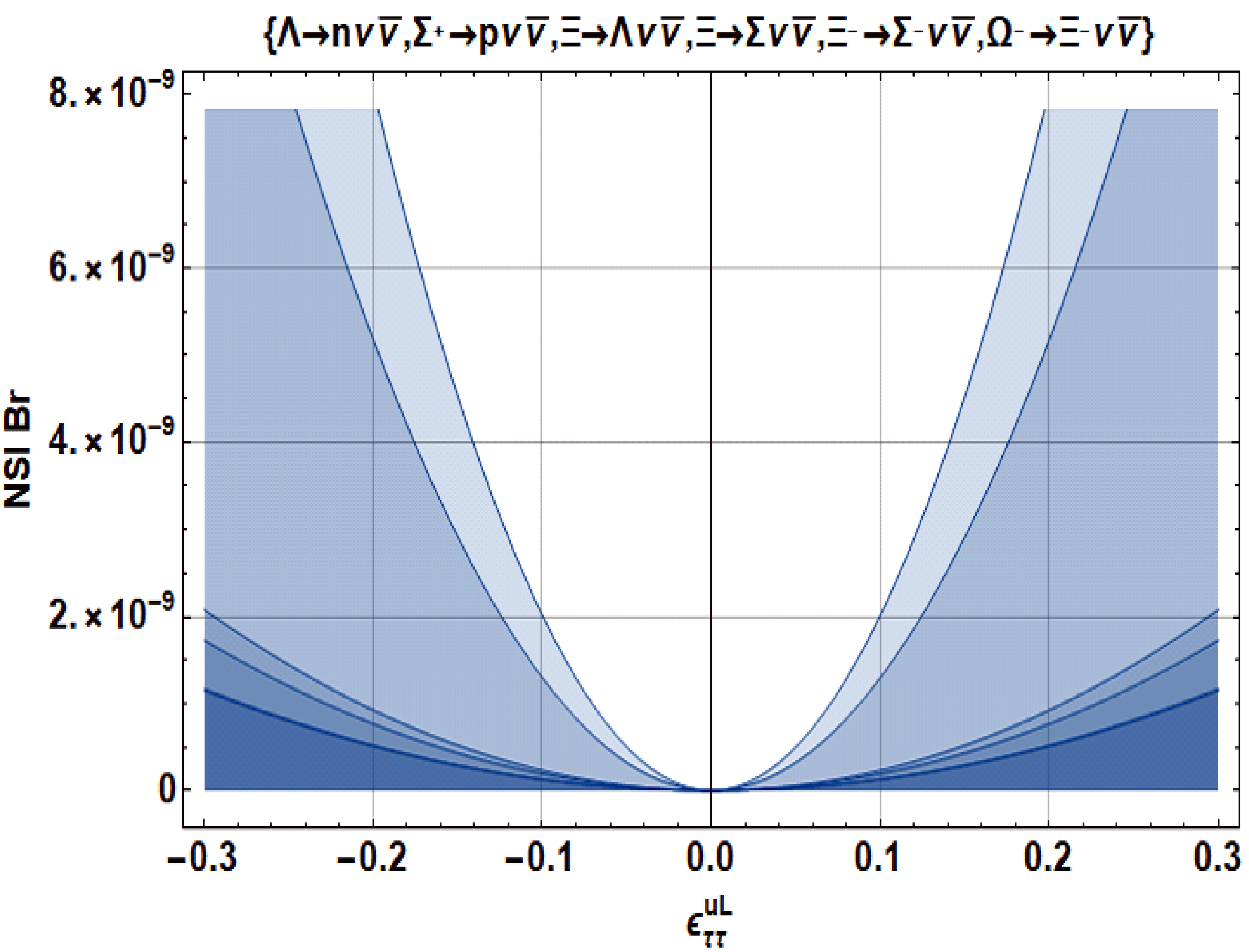

Combine graph of all process with New Physics Parameter $\epsilon_{\tau\tau}$, $O(10^{-2})$

Figure 4: Branching Ratios are very closed to BES-III sensitivity